# Integration demand response aggregator and wind power aggregators with generation units and energy storages in wholesale electricity markets

**Ramin Nourollahi[1*], Rasoul Esmaeilzadeh[1]**

[1]Azarbaijan regional electric company

Ramin.nourollahi@tabrizu.ac.ir, rasoul_zadeh@yahoo.com

*corresponding author

**Abstract**

The integration of various energy resources with wind farms, results in improved market-aware management of wind generation variations. Consequently, the proposed method minimizes the negative impact associated with wind generation deviations on the electricity market and energy systems. This paper integrates the multiple energy resources of smart-grid with wind producer to form a virtual power plant (VRP). One of the primary issues of integrating electricity markets in multi electricity markets such as the Day-Ahead (DHM), intraday (INT), and real time market (RLT) is the participation of a VRP. The proposed VRP formed by integration of the on-site generation (ONG), energy storage system (ENS), wind power aggregator (WPG), and demand response (DRE) contracts. Furthermore, The DRE includes the detailed contracts with shiftable loads and load curtailment load aggreators. A effective method to manage the uncertainty of a VRP in order to attain optimal strategy and offering is provided in this article. The p-robust method is the proposed solution to measure the regret of VRP in the offering strategy's formation. In the modeled scenario-based program, the p-robust approach is used for controlling the relative regret of a VRP in multi-market operation under uncertainty. Obtained results demonstrate that a 6.81% reduction in estimated profit can result in a 45.75% reduction in relative regret of VRP.

## A. Indices

| | |
|---|---|
| $h$ | hour index. |
| $n$ | Contracts providing LOR (Load reduction) index. |
| $s$ | Index of load shifting. |
| $c$ | Index of load curtailment. |
| $e$ | Index of energy storage. |
| $g$ | Index of onsite generation. |

## B. Abbreviation

| | |
|---|---|
| $E$SE | Scheduled energy level of the ENS. |
| WPG | Wind Power |
| INT | Intraday market |
| DHM | Day-ahead market |
| VRP | Virtual Power Plant. |
| EMM | Energy multi-market |
| ENM | Energy market |
| UCY | Uncertainty |
| WIG | Wind generation |
| DRE | Demand response |
| RLT | Real-time market |
| MP | Market price |
| DAG | Demand response aggregator |
| AGG | Aggregator |
| ENS | Energy Storage System method. |
| $OG$ | Onsite generation method. |
| $LS$ | Load shifting method. |
| $SC$ | scheduled energy for generation system. |
| $LC$ | Load curtailment method. |
| BAM | balancing market. |

**C. constant**

| | |
|---|---|
| $\rho_h$ | Energy MP at hour h. |
| $T_n^{min,x}$ | Min load decreasing period of the $n$th Strategy Contract $x$. |
| $T_n^{max,x}$ | Max load decreasing period of the $n$th Strategy Contract $x$. |
| $MN_c^{LC}$ | Max number of daily load curtailment. |
| $RUL_g^{OG}$ | Ramp –up limit of gth OG contract. |
| $RDL_g^{OG}$ | gth OG contract ramp-down restriction. |
| $c_{n,h}^x , Q_{n,h}^x$ | Price and quantity of $n$th contract for strategy $x$ at time $h$. |
| $SD_g^{OG}$ | Competing on the first OG contract's startup costs. |
| $BS_g^{OG}$ | Competing on the OG contract's startup fuel. |
| $PR_g^{min,OG}$ | bottom confine on energy production of gth OG contract. |
| $PR_g^{max,OG}$ | upper confine on energy production of gth OG contract. |
| $MT_g^{on,OG}$ | Indicates the minimum amount on time of gth in the OG contract. |
| $MT_g^{off,OG}$ | Indicates the minimum amount off time of gth in the OG contract. |
| $fC_g^{OG}$ | Indicates the fuel factor gth in the OG contract. |
| $FC_g^{max,OG}$ | High fuel utilization limit in gth OG contract. |
| $EC_e^{max,ES}$ | This symbol represents the energy capacity of the ES facility from the eth of ES contract. |
| $PR_e^{max,ES}$ | ES facility power rating for eth ES contract. |
| $\eta_e^{D,ES}$ | ES facility discharge efficiency for the eth ES contract. |
| $RDL_e^{ES}$ | The symbol for the charging ramp of the ES facility is the ES eth contract. |
| $RD_e^{ES}$ | The symbol for the discharging ramp of the ES facility is the ES eth contract. |
| $ERT_e^{ES}$ | Energy protection time of eth ES contract. |
| $MC_e^{ES}$ | Indicates the maximum allowable charging and discharging times in the ES contract. |
| $T_{e,h}^{ES}$ | Discharging power of $e$th ES contract at time $h$. |
| TN | Represent the sum of scenarios |
| $N$h | Represent the total amount of load in hours |
| $\pi_n$ | Probability of scenario n occurring. |
| $PGw^{Max}$ | Maximum capacity allowed for VRP. |
| $Pc_{Exp/Com}^{Max}$ | Maximum expanding/compressing capacity of ES. |

**D. variables**

| | |
|---|---|
| $\eta^{+/-}$ | This price ratio symbol shows the positive-negative price difference. |
| $\varepsilon h^{+/-}$ | HPP's generating force varies positively or negatively depending on the planned power. |
| $Uc$ | The on-off position for the ES function is represented as a binary variable. |
| $\varepsilon h$ | Binary variable to indicated the ON/OFF stratus of ES performance. |
| $SC_{g,h}^{OG}$ | This symbol represents the cost of startup an OG contract return at time h. |
| $SFC_{g,h}^{OG}$ | Keep gth OG's startup fuel for period h. |
| $T_{g,h}^{OG}$ | gth OG contract producing power at time h. |
| $P_h^x$ | At period h, the total scheduled load of strategy x is decreasing. |
| $CP_n^x$ | At time h, the total cost of the load-decreasing technique x. |
| $PIA_{n,h}^x$ | Load decreasing initiation price of nth hold of x at period h. |
| $F_{n,h}^x$ | nth LOR contract of strategy Xh hours h LOR status indicator; 1 if the contract is scheduled, 0 otherwise. |
| $F_{n,h}^x$ | Refers to the start of the nth contract for the LOR strategy at hour h. |
| $W_{n,h'}^x$ | Refers to stopping the nth contract for a LOR strategy at hour h. |
| $IA_n^x$ | Offered LOR initiation cost of nth contract of strategy x. |

## I. Introduction

*1.1 propellant and purpose*

By definition, renewable energy sources like wind power, might be a realistic option for reducing pollution and greenhouse gas emissions from large-scale power plants. However, due to energy imbalance costs, Wind Power Generators (WPGs) are constrained in their engagement in electricity markets due to energy unbalancing costs [1], [2] . also there are several solutions that support WPGs, such as allocation Subsides or unique price lists to keep them in markets, these solutions are incompatible with competitive marketplace principles; as a result, a market-based strategy is frequently recommended to increase renewables penetration [3], [4]. Corrections made after the Day-ahead market (DHM) gate is closed can dramatically enhance WPG prevalence in electricity market while also lowering the amount and rate of power in RLT [5]. Within the literature, Demand response (DRE) and energy storage systems (ENS) are also used to manage the Uncertainty (UCY) of Wind generation (WIG) and, as a consequence, to boost WPG profitability and penetration in Energy market (ENM). In RLT, ENSs may also be utilized to mitigate energy imbalances [6], [7]. As a result, virtual power plants (VRPs) can provide a vital role in the energy market by addressing the UCY associated with wind generation. Although DRE, ENS, and VRP have all been used in the literature for WIG applications, there is a gap in the literature in terms of modelling a VRP that uses DRE and ENS together to maximize WPG profits in the DHM and INT. This VRP version produces a bigger return for WPG than previous versions, and as a consequence, its higher units set the ground for renewable electricity to reclaim its place in competing ENM.

1.2. Literature review

Utilizing WPGs in diverse electric markets has been studied drastically with inside the literature. For example, in [8] a dihedral modelling technique for strategic offering of a WPG with market power inside the DHM as a price maker and in the balancing market as a deviator is presented. To assess a WPG's ideal supplying strategy for operating in DHM and BAM as a price maker organization, the authors of [2] suggested adopting a multi-level risk-constrained model. Within the literature, the usage of DRE and ENS is also discussed. In [3], the authors suggested a stochastic concept for WPG participation in different ENM, with DRE as an incoherent operation issue. [9] proposes a WPG that interacts with DHM and RLT employing an ENS, with the WPG portrayed as a price-taker within the markets. To reduce wind UCY, a VRP combining WPG and DRE has been proposed. The authors proposed that WPG and a storage unit work together in DHM and INT in [10], [11]. A VRP comprising of WPG and DRE is also proposed to decrease wind UCY. Optimal energy and reserve bids are established, and the SCM is reduced to a convex optimization, to secure the profitability of private ENS investments. To ensure the profitability of particular ENS investments, optimized energy and protection bids are determined and the stochastic model is reduced to an optimization technique. By reducing the stochastic model into internal partitions, [12], [13] investigates the optimum bidding, scheduling, and implementation of battery ENS in the DHM. The authors of [14] investigated DRE trading in DHM using a two consecutive actual price method. Similarly, in the DHM model, [15] has recommended WPG. [16]describes an INT mechanism that takes Real-time market (RLT) related data from WPG and shiftable load information into account. In this case, the WPGs decide depending on the Market price (MP) for several market transactions at different hours. In addition, the authors of [17] evolved a version for the electricity bidding trouble for VRP along with investors with inside the regular ENM and the INT related DRE. Furthermore, the coupon-based totally DRE program is implemented in [18]to cooperate with WPG for the purpose to acquire most suitable marketplace engagement. The DHM model is developed based on DRE abilities for density  management with WPG related UCY was presented in [19] and[20]. Finally, in [21]–[23], the capacity of available resources such as DRE and ENS to minimize wind energy curtailment and virtual bidding while boosting system flexibility has already been shown.

*1.3. Contribution*

A framework for integrating multiple energy resources such as Demand response aggregator, OGN, and ENSs into Energy multi-market (EMM) is presented in this paper. According to the findings, applying the suggested approach compensates for WIG deviations by deploying additional VRP units, in addition to boosting the participation of several VRP units in the MUM. The recommended P-robust approach is also utilized to improve the operator's risk management of the developed scenarios and to select the strategy based on the operator's desired conservation level. This report also includes the offering curves associated with the VRP's engagement in the DHM. The primary distinction in this article strategy is that it models contract factors such market pricing for VRP components like OGN, ENS, WPG, and the DAG. In comparison to prior studies, the proposed model is far more detailed.

The following are the paper's contributions:

1) Providing a model for VRP's possible units such as OGN, WPG, ENS, and DRE options to contribution in the day-ahead, intraday and  balancing markets.

2) Realistic modeling of usints with all the details including contracting and implementation costs of DREs, shut-down and start-up costs of ONG units and degradation of ENSs.

3) Using a scenario-based stochastic optimization approach for modeling VRP uncertainties and comprehensive risk management using the p-robust strategy.

*1.4. Paper structure*

The structure of this article is stated in 6 sections as follows:

The formulation of the VRP components, including the OGN, ENS, WPG, and DRE, is shown in section 2. Additionally, the proposed architecture of VRP involvement in electricity multi-markets is demonstrated in Section 3. In addition, part 4 depicts the simulation of a VRP. Moreover, WIG scenario and MP modeling presented in section 5. Additionally, section 6 shows the stochastic p–robust optimization method. Finally, conclusions of the paper are represented in Section 7.

## 2. Suggested self-scheduling model for proposed VRP components
### 2.1. DAG modeling

The overall goal of an Aggregator (AGG) is to maximize its pay in the DHM. In order to make a profit, the aggregator sells the product to the energy market and pays the hourly LC paying to the participants. Therefore, the difference in the price of DR in customer contracts and the hourly price of the electricity market will be an incentive for the DAG. Finally, the objective function related to the adder is maximized as follows:

$$Maximize \sum_{h \in Nh} \left[ \rho h \left( P_h^{LC} + P_h^{LS} + P_h^{OG} + P_h^{ES} \right) - \left( CP_h^{LC} + CP_h^{LS} + CP_h^{OG} + CP_h^{ES} \right) \right] \quad (1)$$

According to the objective function, the first line shows the income obtained from the total sales of four LOR strategies. Also, the second line refers to the cost paid to the contract customers in order to reduce the load. According to equation 1, the predicted prices in the energy market will be forced to compensate for the LC [24]. AGGs are believed to determine the hourly estimate for the MP of energy, which is known using numerical approaches such as time series and artificial neural networks [25]. In order to estimate hourly MPs, AGGs use past MP data and their DR operating expertise. As a result, the proposed version assumes that the outcomes of several DR AGGs and different marketplace traits on marketplace pricing are already blanketed into the self-scheduling problem`s price forecasting.

### 2.1. Load Curtailment

LC contracts include LC price $C_{n,h}^{LC}$, which is determined by agreements between AGGs and consumers. And also this contract indicates the quantity $Z_{c,h}^{LC}$, which is the AGGs LC of registered clientele in the cth contract. The LCU amount is set up with the aid of using the AGG`s estimation of the DRE functionality of the customers. The LC agreements additionally specify the minimal and most period for LOR, the most range of every day load curtailments, and the price of LOR initiation. The following is the advised method for LC contracts:

$$P_h^{LC} = \sum_{n \in N_{LC}} Z_{c,h}^{LC} Y_{c,h}^{LC} \quad (2)$$

$$CP_h^{LC} = \sum_{n \in N_{LC}} \left( PIA_{c,h}^{LC} + C_{c,h}^{LC} + Z_{c,h}^{LC} + Y_{c,h}^{LC} \right) \quad (3)$$

$$PIA_{c,h}^{LC} \geq IA_c^{LC} F_{c,h}^{LC} \quad \forall c, \forall h \quad (4)$$

$$\sum_{h=h'}^{h+PD_k^{\min,LC}-1} Y_{c,h'}^{LC} \geq PD_c^{\min,LC} F_{c,h}^{LC} \quad \forall c, \forall h \quad (5)$$

$$\sum_{h'=h}^{h+PD_k^{\max,LC}-1} W_{c,h'}^{LC} \geq F_{c,h}^{LC} \quad \forall c, \forall h \quad (6)$$

$$\sum_{h \in H} F_{c,h}^{LC} \leq MC_c^{LC} \quad \forall c \tag{7}$$

$$F_{c,h}^{LC} - W_{c,h}^{LC} = Y_{c,h}^{LC} - Y_{c(h-1)}^{LC} \quad \forall c, \forall h \tag{8}$$

$$F_{c,h}^{LC} + w_{c,h}^{LC} \leq 1 \quad \forall c, \forall h \tag{9}$$

A summed LOR delivered by LC contracts $P_h^{LC}$ at period and the cost function related with them are described $CP_h^{LC}$ in (2)–(3). A binary variable $Y_{c,h}^{LC}$ is connected with the cth LC contracts, which is 1 if the AGG schedules the offer. Function (4), also shows the initial cost of LOR. The binary variable $W_{c,h}^{LC}$ denotes whether or not the LOR in the $m$th arrangement would begin at hour $h$. The constraints on minimal LOR duration, most LOR duration, and the most variety of each day load curtailments are provided in (5) – (7). The contract's start/stop indicators are determined by constraint (8), whereas constraint (9) ensures that such binary variables do not all turn out to be 1 at the same time.

*2.2. Load Shifting*

Constraints (10) – (17) constitute the advised version for LS contracts. The aggregated LOR, the associated cost function, and the LOR initiation price of LS preparations are all covered in the LS version in (10) – (12), as are the restrictions at the minimal and most period of LOR in (13) – (14). Also, ($MT_h^{LS}$, Hs$^{LS}$, α$_n$) determines the three data points for LS contracts, representing that customers will be able to change their load by α$_n$ percent from $H_s^{LS}$ to $H_s^{SH}$ periods under the LS contract. In (17), the AGG ensures that LS arrangements will be available during the period. The AGG would provide the shifting info for the scheduled LS arrangement to the ISO, among other contract information.

$$P_h^{LS} = \sum_{s \in N_{LS}} Z_{s,h}^{LS} Y_{s,h}^{LS} \tag{10}$$

$$CP_h^{LS} = \sum_{s \in N_{LS}} \left( PIA_{s,h}^{LS} + c_{s,h}^{LS} + Z_{s,h}^{LS} + Y_{s,h}^{LS} \right) \tag{11}$$

$$PIA_{s,h}^{LS} \geq IA_s^{LS} F_{s,h}^{LS} \quad \forall s, \forall h \tag{12}$$

$$\sum_{h'=h}^{h+LRD_k^{\min,LS}-1} Y_{s,h'}^{LS} \geq PD_s^{\min,LS} F_{s,h}^{LS} \quad \forall s, \forall h \tag{13}$$

$$\sum_{h=h'}^{h+LRD_k^{\max,LS}-1} W_{s,h'}^{LS} \geq F_{s,h}^{LS} \quad \forall s, \forall h \tag{14}$$

$$F_{s,h}^{LS} - W_{s,h}^{LS} = Y_{s,h}^{LS} - Y_{s(h-1)}^{LS} \quad \forall s, \forall h \tag{15}$$

$$F_{s,h}^{LS} + W_{s,h}^{LS} \leq 1 \quad \forall s, \forall h \tag{16}$$

$$Y_{s,h}^{LS} = 0 \quad \forall h \notin H_s^{LS} \tag{17}$$

2.3. Utilizing Onsite Generation

The recommended method for equivalent LORs given by OSG contracts is represented by requirements (18) - (27). The sum of power produced by clients' onsite generating fleet is used to calculate the overall equivalent LOR delivered under OSG contracts (18). Relationships (19) and (20) represent the start-up cost and the operation cost associated with OSG data. Equation (21) shows the high and low limits on the amount of electricity generated in the OSG contract. equation (22) - (25) also represent the time constraints of increases/decrease and minimum on/off of the production fleet, respectively. According to the text of equation (26), the upper limit of the total fuel consumption of the production fleet is in the OSG contract, and then the amount of fuel required in equation (27) is shown.

$$P_h^{OG} = \sum_{g \in N_{OG}} T_{g,h}^{OG} \tag{18}$$

$$CP_h^{OG} = \sum_{g \in N_{OG}} \left( SC_{g,h}^{OG} + c_{g,h}^{OG} T_{g,h}^{OG} \right) \tag{19}$$

$$SC_{g,h}^{OG} \geq SP_g^{OG} \left( u_{g,h}^{OG} - u_{g(h-1)}^{OG} \right) \quad \forall g, \forall h \tag{20}$$

$$Y_{g,h}^{OG} T_g^{\min,OG} \leq T_{g,h}^{OG} \leq Y_{g,h}^{OG} T_g^{\max,OG} \quad \forall g, \forall h \tag{21}$$

$$T_{g,h}^{OG} - T_{g(h-1)}^{OG} \leq RUL_{g,h}^{OG} \quad \forall g, \forall h \tag{22}$$

$$T_{g(h-1)}^{OG} - T_{g,h}^{OG} \leq RDL_{g,h}^{OG} \quad \forall g, \forall h \tag{23}$$

$$\sum_{h',h}^{h+T_g^{on,OG}-1} Y_{g,h'}^{OG} \geq MT_g^{on,OG} \left( Y_{g,h}^{OG} - Y_{g(h-1)}^{OG} \right) \quad \forall g, \forall h \tag{24}$$

$$\sum_{h'=h}^{h+T_g^{off,OG}-1} \left( 1 - Y_{g,h'}^{OG} \right) \geq MT_g^{off,OG} \left( Y_{g(h-1)}^{OG} - Y_{g,h}^{OG} \right) \quad \forall g, \forall h \tag{25}$$

$$\sum_{h \in H} \left( T_g^{OG} T_{g,h}^{OG} + SFC_{g,h}^{OG} \right) \leq T_g^{\max,OG} \quad \forall g, \forall h \tag{26}$$

$$SFC_{g,h}^{OG} \geq BS_g^{OG} \left( Y_{g,h}^{OG} - Y_{g(h-1)}^{OG} \right) \quad \forall g, \forall h \tag{27}$$

2.4. Utilizing ENS System

The capabilities of the ENS fleet of customers are laid out in an ES agreement. The proposed ES contract model is shown in (28) – (37). The total the scheduled LOR and its price expressed through ENS contracts is the general LOR added through the AGG via ES contracts and the associated price characteristic at time (h), denoted through (28) – (29). Equation (30) limits the total LOR to less than power, while equation (31) and (32) limit the ENS rating between two consecutive hours. It can be seen from equation (33) that the AGG ensures that the energy capacity of the ENS fleet is greater than the total amount of ENS capacity during the scheduled hours in a contract, even taking into account the discharge efficiency. Equation (34) specifies the minimum daily allowable level for charging and discharging. Equation (35) shows the energy conservation constraint at the eth of the ES contracts. With the research done in reference 22, it can be claimed that in equation (35), the storage time of energy in an ENS fleet will not be longer than the storage time[26]. Relationships (36) and (37) also represent binary variables.

$$P_h^{ES} = \sum_{e \in N_{ES}} T_{e,h}^{ES} \tag{28}$$

$$CP_h^{ES} = \sum_{e \in N_{ES}} c_{e,h}^{ES} T_{e,h}^{ES} \tag{29}$$

$$0 \leq T_{e,h}^{ES} \leq Y_{e,h}^{ES} T_e^{\max,ES} \quad \forall e, \forall h \tag{30}$$

$$T_{e,h}^{ES} - T_{e(h-1)}^{ES} \leq RUL_{e,h}^{ES} \quad \forall e, \forall h \tag{31}$$

$$PD_{e(h-1)}^{ES} - PD_{e,h}^{ES} \leq RDL_{e,h}^{ES} \quad \forall e, \forall h \tag{32}$$

$$\sum_{h \in H} PD_{e,h}^{ES} \leq \eta_e^{D,ES} EC_e^{\max,ES} \quad \forall e \tag{33}$$

$$\sum_{h \in H} F_{e,h}^{ES} \leq MC_e^{ES} \quad \forall e \tag{34}$$

$$\sum_{h'=h}^{h+ERT_k^{ES}-1} W_{e,h'}^{ES} \geq F_{e,h}^{ES} \quad \forall e, \forall h \tag{35}$$

$$F_{e,h}^{ES} - W_{e,h}^{ES} = Y_{e,h}^{ES} - Y_{e(h-1)}^{ES} \quad \forall e, \forall h \tag{36}$$

$$F_{e,h}^{ES} + W_{e,h}^{ES} \leq 1 \quad \forall e, \forall h \tag{37}$$

In the described method, (1) - (37) represent the DAG's self-scheduling model in the DHM. The recommended technique is written as a mixed-integer linear programming (MILP) hassle that may be solved the use of any MILP solver available. For a specific marketplace pricing profile, the model`s conclusions set up the best participation offering of a DAG with inside the DHM. For LOR contracts, the best DRE offering consists of each the hourly price and the quantity. Also, the reduction of the load resulting from the LS and ES contracts has created periods for shifting loads and charging the ES fleet. During the hours whilst the predicted price of power is low, affordable DAGs might charge the ENS and recoup the decrease load in LS contracts. This may bring about an unexpected growth in load all through certain hours, complicating the ISO's functioning or even posing a safety danger to the system The ISO might deal with ENS structures as aggregated negative load contracts in our proposed DAG framework, co-optimizing the charging intervals of ENS and the shifting duration of LS contracts all through the DHM scheduling horizon, whilst scheduling LORs and clearing the DHM. As a result, the local DRE could be concerned with inside the system performance

## 3. Virtual Power Plant (VRP) Modeling

This phase depicts a VRP bidding method for maximizing take advantage of the blended operation of a WPG, a DAG, and an ENS. Figure 1 depicts the participation of VRP along with the ONG, ENS, WPG, and DRE in the EMM, which includes the DH, INT, and BM. This section depicts an VRP bidding technique for maximizing profit from the combined operation of a WPG, a DAG, and an ENS. The OF for VRP's participation in all ENM is as follows:

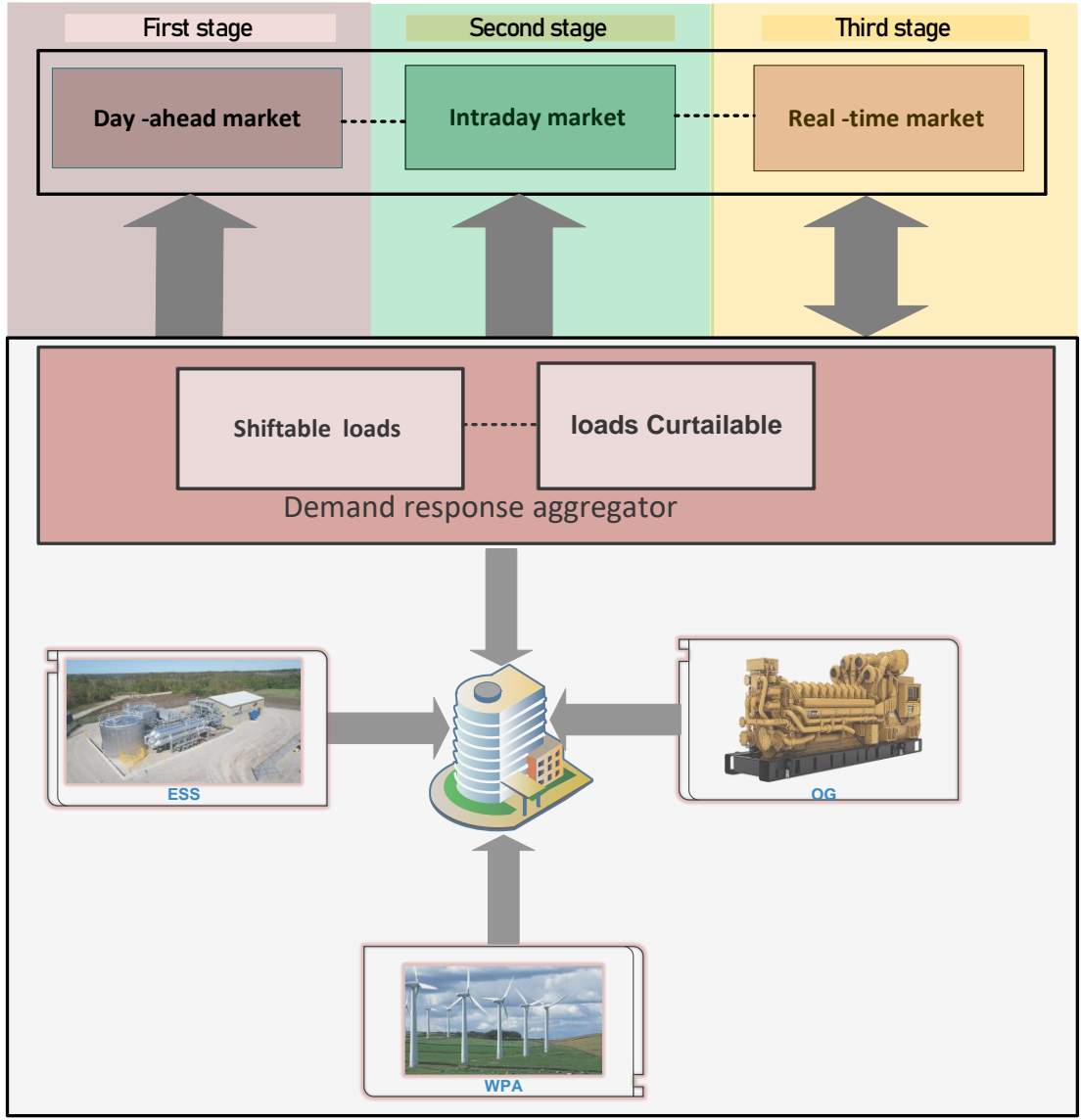

Fig.1. the VRP participation in the multi-market

$$Max_{\Theta H_{n,h};\varphi_n,\forall n;\theta} \left[ Z_{profit}^{VRP} \right]$$
$$= \sum_{n=1}^{N_n} \sum_{h=1}^{N_h} \pi_n [\rho_{n,h}^{DA} . PG_{n,h}^{DA} + \rho_{n,h}^{IN} . PG_{n,h}^{IN} \quad (38)$$
$$+ \rho_{n,h}^{DA} . \eta_{n,h}^{+} . \varepsilon h_{n,h}^{+} - \rho_{n,h}^{DA} . \eta_{n,h}^{-} . \varepsilon h_{n,h}^{-} ]$$

According to the relationship $\Theta h_{n,h} = \{PG_{n,h}^{DA}, PG_{n,h}^{IN}, \varepsilon h_{n,h}^{+}, \varepsilon h_{n,h}^{-} \; \forall n, h\}$ are the variables related to the VRP optimization problem. the objective function related to equation (38) shows the expected profit of VRP, which is equal to the result of market transaction and operation cost. VRP income is obtained from the purchase and sale of energy in the DHM and INT markets, as well as the cost of energy deviations in the BAM. In (38), the terms $\rho_{n,h}^{DA} . Ph_{n,h}^{DA}$ and $\rho_{n,h}^{IN} . Ph_{n,h}^{IN}$ involve to the revenue/cost of VRP from the DHM and INT markets, while the terms $\rho_{n,h}^{DA} . \eta_{n,h}^{+} . \varepsilon h_{n,h}^{+}$ and $\rho_{n,h}^{DA} . \eta_{n,h}^{-} . \varepsilon h_{n,h}^{-}$ propose the revenue/cost through positive/terrible power deviations with inside the RLT. The VRP optimization problem in objective function

(38) is constrained by some combined constraints that apply to all DAG, ENS, and WPG modelling, as well as some constraints that apply only to DAG, ENS, and WPG modelling. In order to function as an VRP, the following constraints must be met by the DAG, ENS, and WPG:

$$PGH_{h,n}^{\beta} = PGW_{h,n}^{\beta} + PGc_{h,n}^{\beta} \quad \forall h,n \quad \beta = DA, SC, IN \tag{39}$$

$$PGh_{h,n}^{SC} = PGh_{h,n}^{DA} + PGh_{h,n}^{IN} \quad \forall h,n \tag{40}$$

$$\varepsilon h_{h,n} = PGw_{h,n}^{Re} + PDc_{h,n}^{Re} - PGh_{h,n}^{SC} \quad \forall h,n \tag{41}$$

$$\varepsilon h_{h,n} = \varepsilon h_{h,n}^{+} - \varepsilon h_{h,n}^{-} \quad \forall h,n \tag{42}$$

$$0 \leq \varepsilon h_{h,n}^{+} \leq PGw_{h,n}^{Re} + PGc_{h,n}^{Re} \quad \forall h,n \tag{43}$$

$$0 \leq \varepsilon h_{h,n}^{-} \leq PGw^{Max} + PGc_{EXP}^{Max} \quad \forall h,n \tag{44}$$

$$\left(PGh_{h,n}^{DA} - PGh_{h,n'}^{DA}\right).\left(\rho_{h,n}^{DA} - \rho_{h,n'}^{DA}\right) \geq 0 \quad \forall h,n,n' \tag{45}$$

$$PGh_{h,n}^{DA} = PGh_{h,n'}^{DA} \quad \forall h,n,n' \ : \ \rho_{h,n}^{DA} - \rho_{h,n'}^{DA} \tag{46}$$

(46) DHM and INT offers, in addition to VRP`s general scheduled energy, are limited by constraint (39). As calculated in, VRP's overall scheduled energy need to be identical to its DHM and INT offers (40). In the BAM period, the total positive and negative imbalances primarily based on VRP scheduled energy and actual wind generation are denoted in (41) to (42). (44). Constraints (45) and (46) are applied which will provide non-decreasing DHM offerings.

## 4. Wind Generation and MPs

The following is modeling of the wind generation and MP uncertainties: For WIG, DHM, INT, and BAM prices, N1, N2, N3, and N4 scenarios are produced. The UCYs are divided as follows: 1) independent uncertainty parameters, such as WIG denoted as (Pw(w)) and DHM price denoted as (DA(d)), and 2) INT price ($\rho^{IN}(d, i)$) Cases that are feasible for each potential investigation of DHM price scenarios. To put it another way, scenarios related to the ID market are derived from scenarios related to the DHM. For all scenarios, the correlation between these stochastic variables is defined as $(\rho^{DA} - \rho^{IN})$ and ($\eta^{+} + \eta^{-} - 1$) due to the dependence of INT and BAM costs on wind electricity generation and DHM price. In addition, the symmetric scenario tree is used to construct the NS = N1 N2 N3 N4 scenarios from the independent and dependent scenarios.

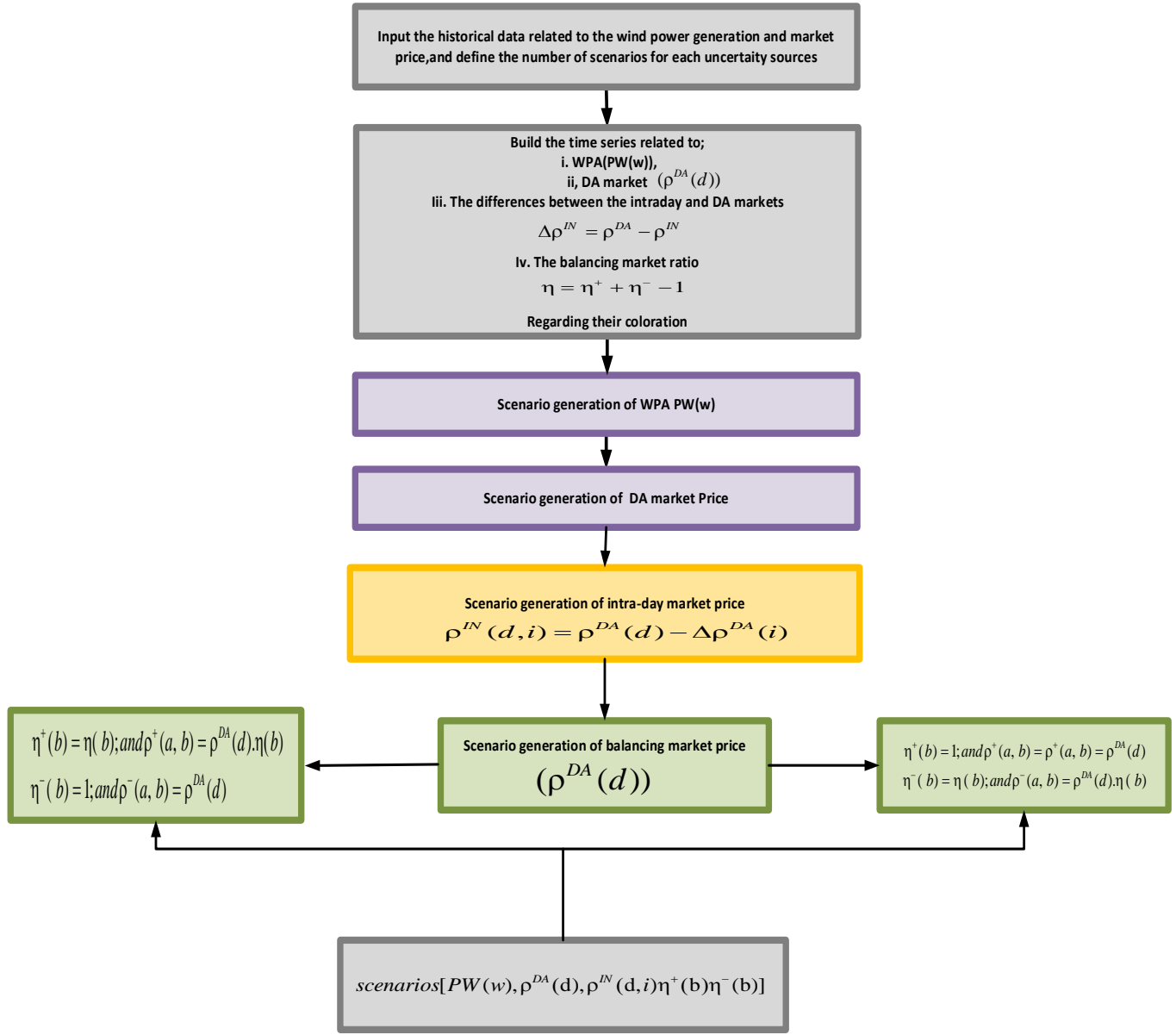

Fig 2. Flowchart related to stochastic variable model

Figure 2 depicts the flowchart for the stochastic modeling that results in the generation of all scenarios $(PW(W), \rho^{DA}(d), (\rho^{IN}(d,i), (\eta^+(b)\eta^-(b))$. It's worth noting that the purple boxes portion reflects the first category of UCY (WIG and DHM pricing possibilities), whereas the yellow and green boxes parts represent the second (i.e., the INT and BAM price scenarios). The highlighted green color portion of Fig2 following scenario production of BAM prices (b) demonstrates that there are two methods for determining balancing MPs based on the values of (b)[27] .

In this part, an optimization method called p-robust is presented, which is used in modeling the scheduling problem of micro grids. in order to avoid uncertainties, the p-robust method includes the advantages of both RO and SO**.** When the usage of the SO approach, the purpose is normally to minimize the predicted cost or maximize the predicted income while taking into consideration all viable scenarios. Although the SO approach`s acquired result is financially best in most of the scenarios, it could bring about losses in others. The RO approach, on the other hand, normally pursuits for the minimum viable cost or the minimum viable regret, making sure that the acquired answer is best for any uncertainties in predefined sets. Nonetheless, it's overly careful due to the fact the result protects in opposition to scenarios which might not be likely to happen. As a result, the stochastic p-robust optimization approach mixes the advantages of each the max-predicted-income and min-max regret optimization techniques through trying to find the p-robust maximal, i.e., in which relative regret in any scenario is not more than p, for P≤0 [28].

*5.1 Method description*

Assume that S represent a number of potential scenarios, And Ps represents a deterministic maximization complex for each scenario $f \in F$. Furthermore, assume $Z^*_f$ be the best solution to problem Ps. It's worth noting that each Ps has the same structure but different data. Let $Z_f(X)$ be the objective value of X in scenarios, and (X) x is a feasible vector for the variables of the model. According to desired robustness level, i.e., $p \leq 0$, X is called p-robust solution if it satisfies the following equation for all f in F:

$$\frac{Z^*_f - Z_f(X)}{Z_f^*} < P; \forall f \qquad (47)$$

The relative regret is demonstrated on the left part of (7). In the same way, $Z^*_n - Z_n(X)$ defines absolute regret. It should be noted that the relative regret is taken into account in the stochastic p-robust optimization model, however this can be modified to consist of the absolute regret. Equation (7) may be rewritten in the following way:

$$Zf(X) \geq (1-p)Z^*_f; \ \forall f \qquad (48)$$

Equation (8) denotes the stochastic p-robustness condition, which could be obtained by combining it with the max predictable profit OF to produce the stochastic p-robust optimization model as follows:

$$\text{Maximize} \sum_{f=1}^{N_f} Q_f Z_f(X) \qquad (49)$$

Subject to:

$$Z_f(X) \geq (1-p)Z^*_f; \ \forall f \qquad (50)$$
$$X \in \chi$$

where $Qf$ indicates the likelihood that scenario s will occur, and c denotes the set of solutions which are possible for all $f \in F$. The stochastic p-robustness condition is enforced by Equation (50). It means that under any solution, the objective value of each scenario $f \in F$ should be greater than $(1-p) Z^*f$. These restrictions are imposed the OF to find a resolve which is both possible across all cases and profitable.

*5.2 Determine upper and lower bounds for the objective function*

Eq. (50) ensures that each scenario's gain is equal to (1-p) Z*s. As a result, the OF exceeds expected overall profit minimum. As a result, the OF's lower bound can be written as represented in the following:

$$LB = \sum_{f=1}^{N_f} Q_f \left(1 - p_f\right) Z_f^* \tag{52}$$

For each scenario, the following equation is valid according to the optimality definition:

$$Zf \geq Zf(X); \quad \forall f \tag{53}$$

Where $Z^*f$ and $Zf(X)$ are, respectively, the best and feasible values related to the OF. In the probability of scenario f, multiply both sides of (53):

$$Q_f Z_f^* \geq Q_f Z_f(X); \quad \forall f \tag{5}$$

If both sides of equation (54) are summed on f, the upper bound for the OF is defined as equation (55):

$$UB = \sum_{f=1}^{Nf} Q_f Z_f^* \geq \sum_{f=1}^{Nf} Q_f Zf(X) \tag{55}$$

## 6. Simulation Results

This section models the coordinated operation of a VRP, which includes the ONG, WPG, DAG, shiftable load and load curtailment, and ENS participation in the multi-market, which includes the four markets. The suggested technique uses stochastic optimization to simulate the VRP and energy market uncertainty as a result of impact of the uncertainties on the electricity market. The scenario creation process is described in the previous section. Figure 3 depicts the possibilities relating to wind power and MP input data. In addition to, tables 3-5 represent the features of the ONG, ESE, and DRE.

Table 3. the characteristics of the OG

| Contract | Min/Max Power (MW) | Price ($/MW) | Startup Cost ($) | Min on/off time (h) | Ramp Up/Down (MW/h) | Startup Fuel (MBtu) | Fuel limit (MBtu) |
|---|---|---|---|---|---|---|---|
| 1 | 1/10 | 45 | 100 | 1 | 10 | 20 | 100 |
| 2 | 1/10 | 45 | 100 | 1 | 10 | 20 | 100 |
| 3 | 1/10 | 50 | 100 | 1 | 10 | 20 | 100 |

Table 4. the characteristics of the ONG, ESE, and DRE

| Contract | Power Rating (MW) | Energy Capacity (MWh) | Efficiency of Discharge | Discharge Ramp (MW/h) | Time to Retain Energy (h) |
|---|---|---|---|---|---|
| 1 | 10 | 60 | 0.9 | 20 | 12 |
| 2 | 10 | 60 | 0.9 | 20 | 12 |
| 3 | 10 | 60 | 0.9 | 20 | 12 |

Table 5. the characteristics of the ONG, ENS, and DRE

| Contract | $T_k^{LSH}$ | $T_k^{SH}$ | $\alpha_k$ |
|---|---|---|---|
| 1 | 10-16 | 4-10 | %100 |
| 2 | 14-20 | 8-1 | %100 |
| 3 | 16-22 | 10-16 | %100 |

Table 6. the characteristics of the ONG, ESE, and DRE

| Contract | Quantity (MW) | Cost of Initiating a LOR ($) | Minimum LOR Time (h) | Maximum LOR Duration (h) | Number of Daily Curtailments Maximum |
|---|---|---|---|---|---|
| 1 | 10 | 100 | 3 | 6 | 1 |
| 2 | 10 | 100 | 3 | 6 | 1 |
| 3 | 10 | 100 | 3 | 6 | 1 |

The outcomes of the VRP improvement model for different values of robustness parameter p are reported in this section. Stochastic optimization is employed in the suggested technique to represent market rate uncertainty. The SPRO is then suggested to demonstrate the financial risks related to uncertainty parameters in the worse-case scenario. The outcomes of the stochastic optimization expect a profit loss for VRP in this case. To demonstrate the approach outcomes, two cases are examined in the suggested SPRO model in comparison with each other. The outcomes of the basic stochastic approach are shown in Case 1 by calculating the model for p=+.

More findings are obtained in Case 2 for the initial quantity for p, which is p= P=0.1314. The obtained findings for p=0.0747 are robustness outcomes for the unknown parameter uncertainty, which is referred to in the literature as a risk-averse approach. Table 1 depicts the scenario-dependent profits of VRP for different quantities of p. Table 1 shows that reducing the p-robust factor lowers VRP profit. The last iteration in p=0.0747, which is the final feasible result, corresponds to a risk-averse approach, whereas the first iteration in p=+ corresponds to a risk-neutral strategy. Table 1 shows that switching from a risk-neutral approach to a risk-averse strategy decreases profit in all scenarios. Figure 2 shows the profit decrement against p variations for all scenarios. Figure 2 depicts the profit loss caused by the reduction of p.

Figure 3 depicts the profit fluctuation in all scenarios for the three introduced cases. Based on Fig. 3, the profit related to p=+ is the highest, while p=0.0747 is the lowest profit that a VRP owner could get in an uncertain situation. A noticeable finding of Fig. 3 is that the profit fluctuation from p=+ to p=0.1314is minor, but the profit variation from p=0.1314 to p=0.0747 is considerable. Based on Fig. 3, the upper and lower profit margins are close together in certain scenarios, such as Scenarios 5, 6, and 7. As a result, under the mentioned circumstances, the DHM price related UCY has several impacts on the VRP profit.

Table 1. Estimated profit variation compared to a reduction of p in all scenarios

Table 2 shows how the expected profit and highest relative regret vary when p is reduced. Table

| P | S1 | S2 | S3 | S4 | S5 | S6 | S7 | S8 | S9 | S10 |
|---|---|---|---|---|---|---|---|---|---|---|
| P=+∞ | 25342.1 | 25368.8 | 25617.8 | 27603.5 | 26338.8 | 27960.6 | 26938.8 | 25607.7 | 28569.9 | 22586.6 |
| P=0.1314 | 25324.1 | 25272.7 | 25390.4 | 27456.5 | 26187.5 | 27004.9 | 26838.3 | 25407.2 | 28566.6 | 22586.6 |
| P=0.1251 | 25346.6 | 24757.0 | 25322.2 | 26923.5 | 26167.9 | 26444.9 | 26823.2 | 25235.3 | 28555.2 | 22586.6 |
| P=0.1188 | 25281.0 | 24282.1 | 25352.7 | 26791.4 | 26101.6 | 25920.1 | 26743.6 | 24770.4 | 28487.4 | 22586.6 |
| P=0.1125 | 23167.0 | 24936.5 | 25166.9 | 27330.6 | 25706.2 | 25711.3 | 26429.0 | 25083.7 | 28379.4 | 22586.6 |
| P=0.1062 | 24719.0 | 24872.4 | 24690.4 | 26838.8 | 25402.4 | 25460.1 | 26200.1 | 24444.8 | 27488.6 | 22586.6 |
| P=0.0999 | 24548.4 | 24309.2 | 24649.6 | 27005.2 | 25209.9 | 25474.5 | 26260.8 | 23942.6 | 26947.9 | 22586.6 |
| P=0.0936 | 24765.3 | 22994.3 | 24367.9 | 25796.9 | 24752.2 | 25396.4 | 26185.9 | 24483.9 | 27861.0 | 22586.6 |
| P=0.0873 | 24646.9 | 23154.1 | 23561.2 | 25420.9 | 24458.2 | 25654.5 | 26248.6 | 24343.8 | 27395.5 | 22586.6 |
| P=0.081 | 24272.4 | 23313.9 | 23789.1 | 25571.5 | 24205.8 | 25695.9 | 25758.7 | 23684.1 | 26895.8 | 22586.6 |
| P=0.0747 | 23493.5 | 23473.8 | 23704.1 | 25541.6 | 24371.8 | 25872.1 | 24926.5 | 23694.8 | 26435.9 | 22586.6 |
| P<0.0747 | Infeasible | | | | | | | | | |

2 shows that by lowering p, the profit and MRR of VRP are lowered. The final row of Table 2 demonstrates VRP's robust technique for managing DHM price related UCY. As a result, Table 2 shows that despite a decrease in earnings, the VRP can be maintained against fluctuations in the DHM price. A noticeable outcome is the lowering of MRR in comparison to the estimated profit drop, as seen in Table 2 and Fig. 4. Table 2 and Fig. 4 demonstrate that MRR and estimated profit in p=+ are 13.77% and $263.47, respectively. while in p=0.0747 they are reduced by 7.47% and $24410.06, respectively. As a result, the VRP's estimated profit and MRR are decreased by 6.81% and 45.75%, from p=+ to p=0.0747. Thus, it is evident that from p=+ to p=0.0747, a 6.81% reduction in estimated profit can result in a 45.75% reduction in MRR, as shown in Fig. 4. It's worth mentioning that p=+ and p=0.0747 correlate to the VRP risk-neutral and risk-averse strategies, respectively. Finally, there are no feasible results for P less than 0.0747.

Table 2. Estimated income compared to MRR through different quantities of p

| P | MRR (%) | MRR reduction (%) | Expected income ($) | Reduce expected revenue(%) |
|---|---|---|---|---|
| P=+∞ | 13.77 | 0 | 26193.47 | 0 |
| P=0.1314 | 13.14 | 4.57 | 26003.49 | 0.73 |
| P=0.1251 | 12.51 | 9.15 | 25816.24 | 1.44 |
| P=0.1188 | 11.88 | 13.72 | 25631.67 | 2.14 |
| P=0.1125 | 11.25 | 18.3 | 25449.72 | 2.84 |
| P=0.1062 | 10.62 | 22.87 | 25270.34 | 3.52 |
| P=0.0999 | 9.99 | 27.45 | 25093.47 | 4.2 |
| P=0.0936 | 9.36 | 32.02 | 24919.05 | 4.87 |
| P=0.0873 | 8.73 | 36.6 | 24747.05 | 5.52 |
| P=0.081 | 8.1 | 41.17 | 24577.4 | 6.17 |
| P=0.0747 | 7.47 | 45.75 | 24410.06 | 6.81 |
| P<0.0747 | Infeasible | | | |

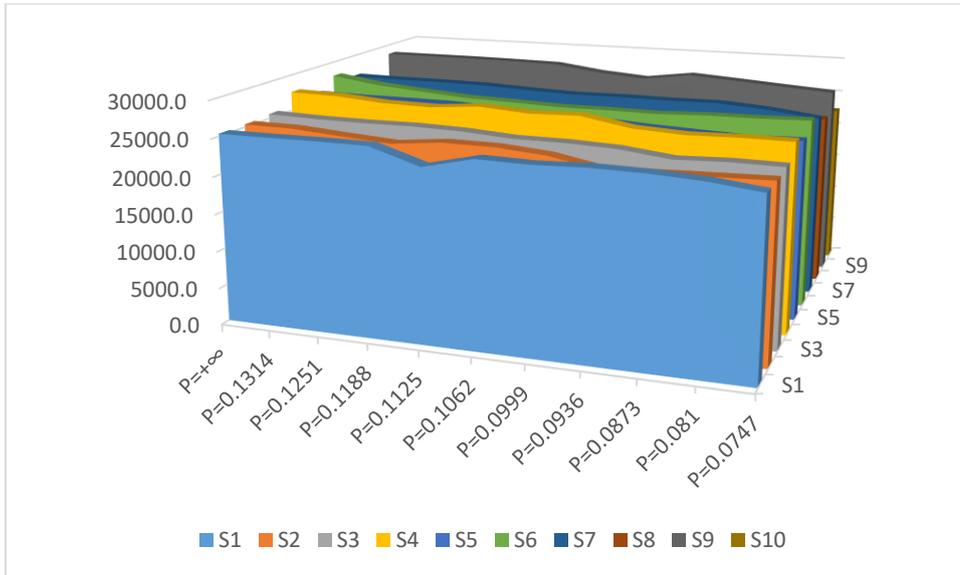

Fig 3. In all of the possibilities, the estimated revenue was compared to a declining p.

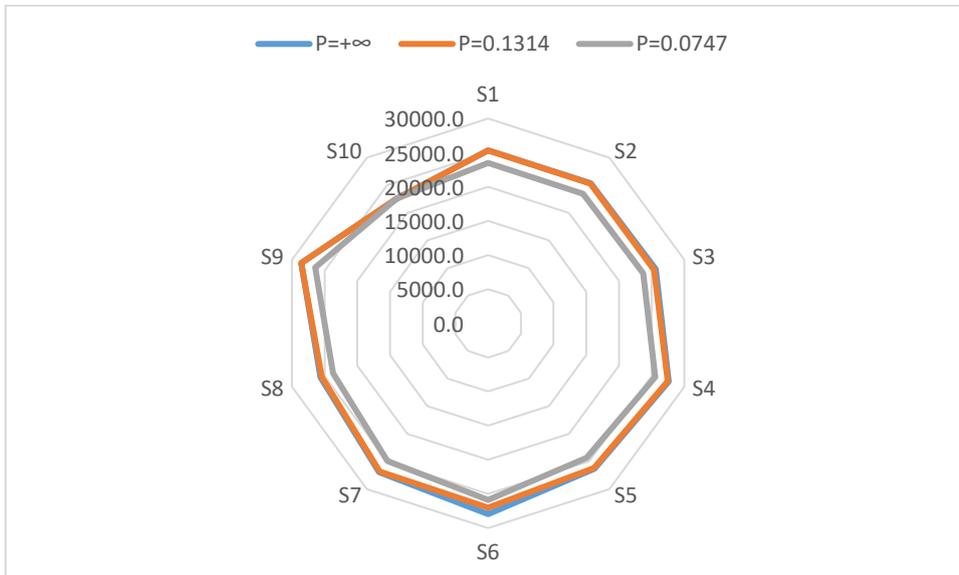

Fig 4. Profit fluctuation for the maximum and minimum limitations in all situations

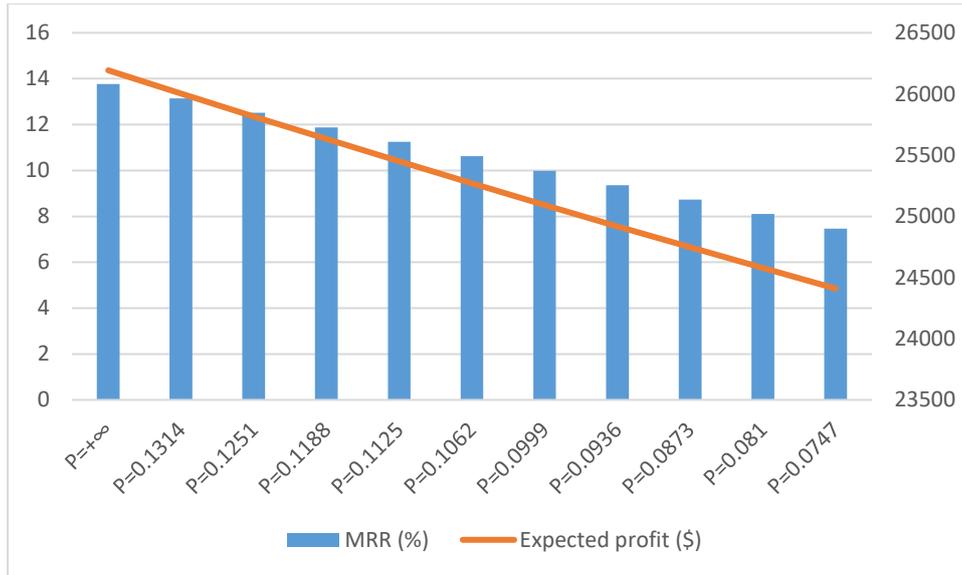

Fig 5. Estimated profits are compared against MRR using a variety of p values.

Fig 6. shows the participation of DRE in the DHM by implementing the RA and RN strategies. According to the analysis, between hours 8 to 13 and 18 to 24, the DRE participation in the DHM through the Risk-neutral Risk-averse methods is considerably higher compared to the other hours of the day due to high MPs. the Risk-averse strategy is considered a conservative strategy. A risk averse strategy is an investor who prefers lower profits with low risks rather than higher profits with high risks. In other words, among various investments giving the same return with different level of risks, this investor always prefers the alternative with least interest. the RA models the worst-case scenario approach. The worst-case scenario approach for the energy sellers models the lowest MP. Thus, the DRE participation in the market through the RA strategy is lower compared to the RN method due to low MPs.

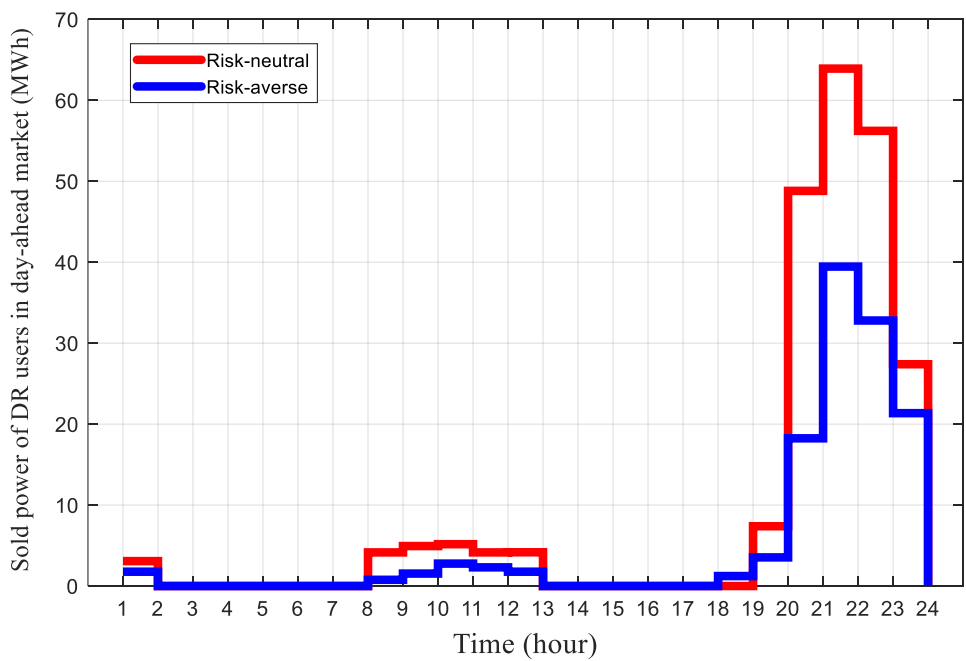

Fig 6. DRE participation in the DHM

The DRE participation in the INT is represented in Fig 7. According to the analysis, it can be seen that Risk-neutral participation between 8 to 11 and 21 to 22 is more than Risk-averse, and the reason by definition can be the conservatism of this strategy. Also, between 23 and 24 hours, it is observed that Risk-averse participation is more than Risk-neutral, which can be due to the participation of other units, especially the wind unit.

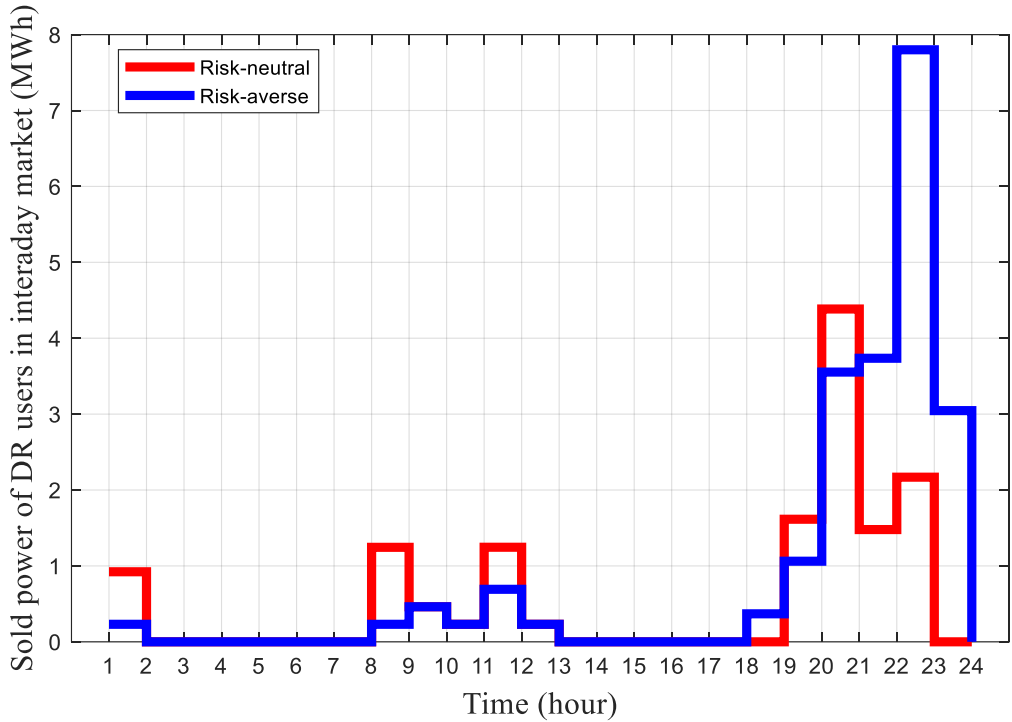

Fig 7. DRE participation in the INT

The following display the ONG`s participation with inside the DAH. The ONG participation with inside the Risk –averse method is substantially low because of the start-up and the working expenses of the ONG devices. Moreover, ONG devices are touchy to price fluctuations that can give an explanation for the large variant among ONG participation via the Risk –averse and the Risk- impartial techniques.

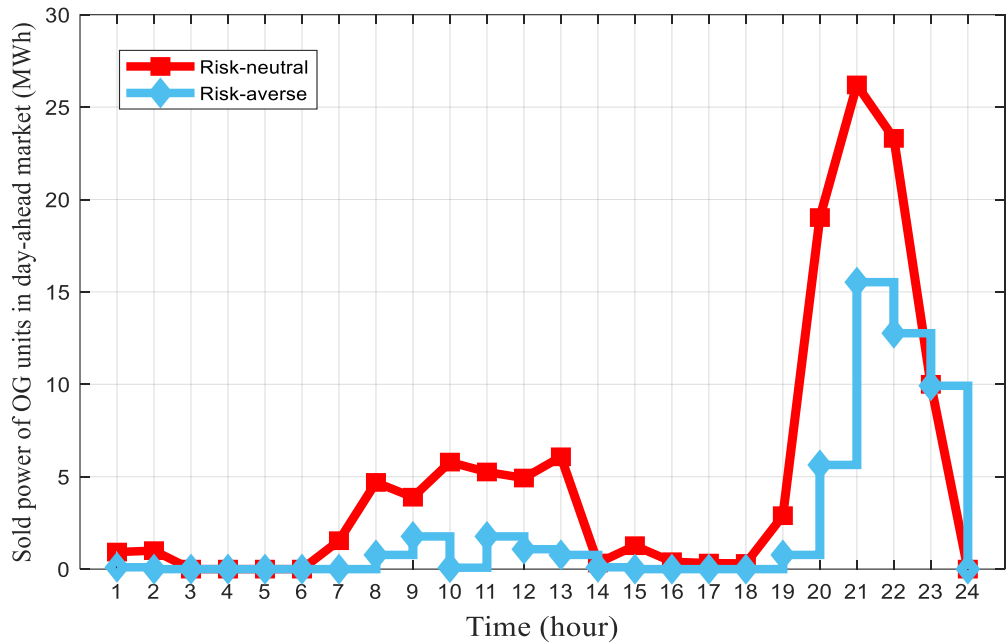

Fig 8. OG's participation in the DHM

Fig 9. refers to the participation of ONGs in the INT. According to the analysis, in the period between 2 and 3, the risk-neutral strategy had less participation in the market, and then between 7 and 15 hours, fluctuations in the risk-neutral strategy are observed due to price changes in this period, operating costs and costs. Is set up. Also, in this curve, it is noticeable that ONGs are significantly dependent on price changes, which in times of price reduction, the market share of these units also decreases.

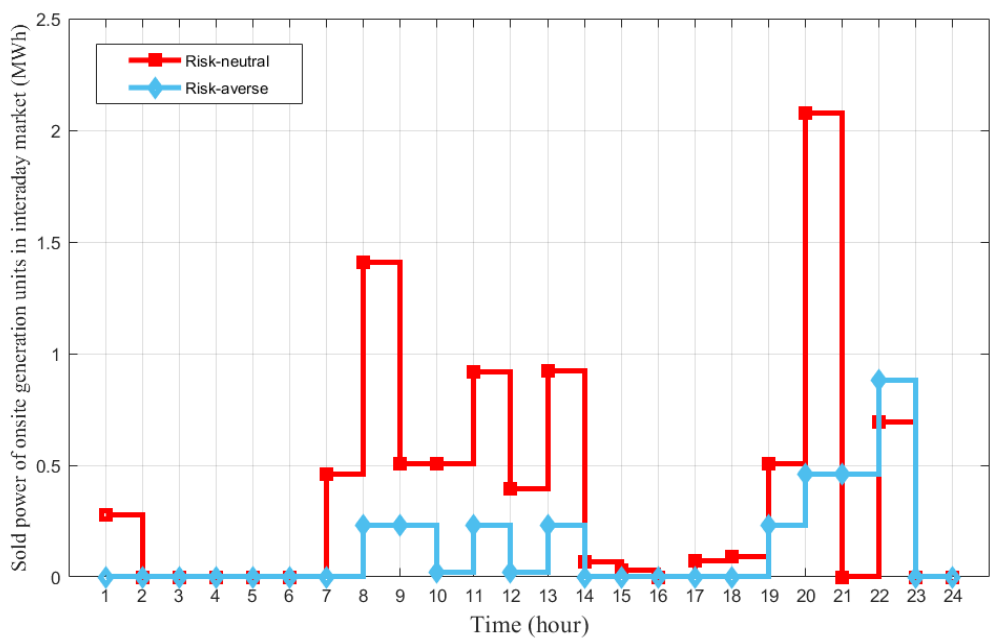

Fig 9. ONG's participation in the INT.

The INT participation in storage units is shown in Figure 10. In this part, Risk-neutral strategy has participated in the market in the period of 1 to 2, but there is no participation from Risk-averse strategy in about 7 hours, which is due to the conservative nature of this strategy. The main difference between these two strategies is this. Risk -neutral power levels at different prices It sells for almost the same amount, but Risk -averse saves most of its power for a high price due to its conservatism.

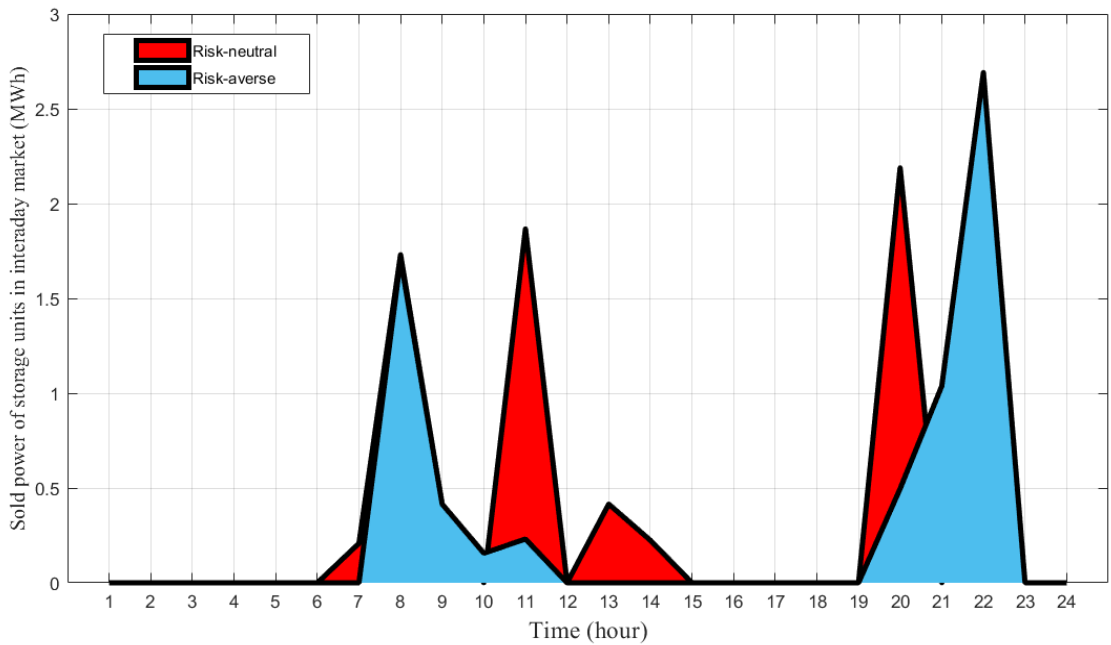

Fig 10. storage unit participation in the INT

Fig 11. depicts the participation in storage unit in the DHM. In this curve, the Risk-neutral and Risk-averse strategies sell equal amounts of their power, then the Risk-averse strategy stores a large amount of its power to sell at the high peak, but the Risk-neutral strategy in the range of 10 to 12 Also sells as much as the previous interval and because it tends to one High profit trading sells its power at different intervals.

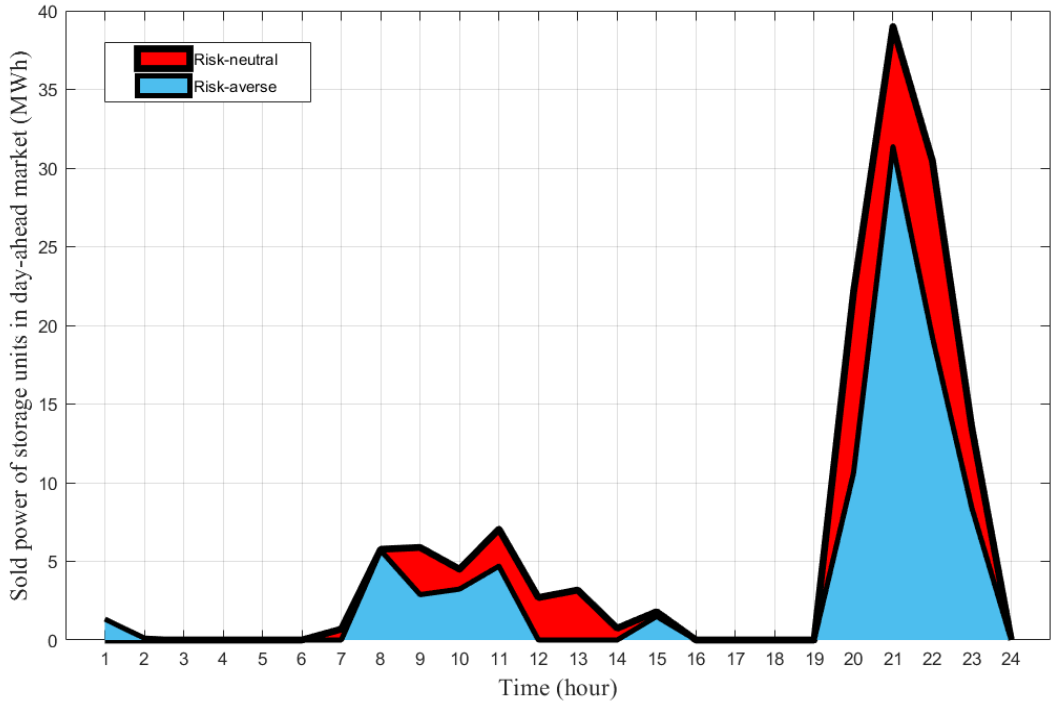

Fig 11. storage unit participation in the DHM

As can be seen, Figure 12 refers to participation in the INT and DHM. According to the analysis performed in this curve, the participation of Risk-neutral strategy was higher than Risk-averse, but the difference between the two strategies in this curve is much less compared to the-previous curves, due to price fluctuations in different time periods.

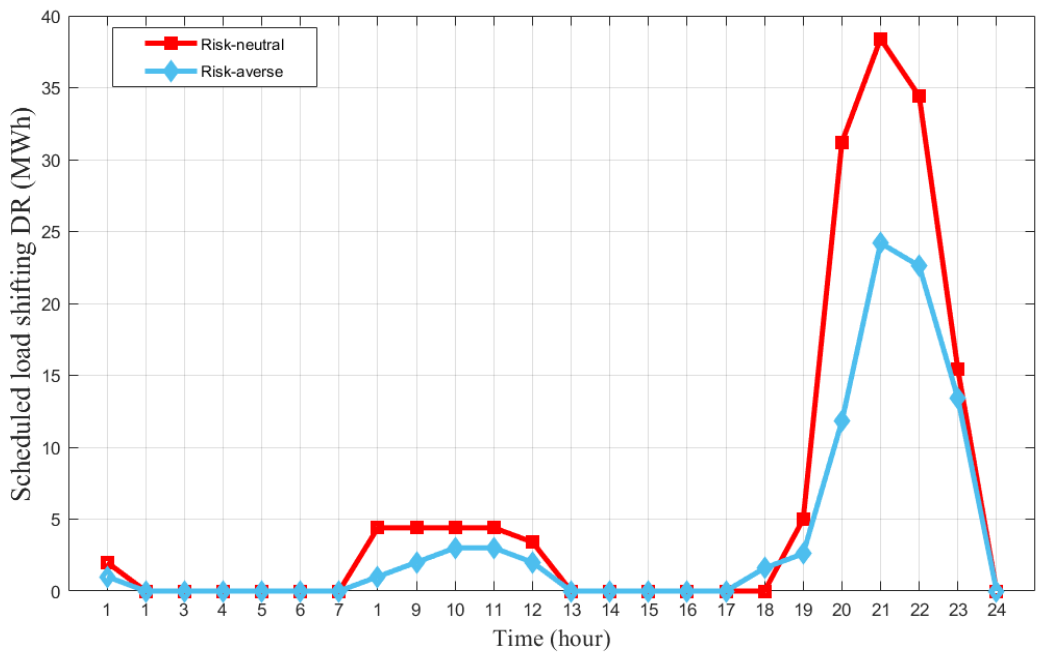

Fig 12. DR participation in the INT and DHMs

Fig 13. represents the scheduled load curtailment related to the DRE in the DHM and INT. According to the analysis in the time range of 1 to 3, both Risk-averse and Risk-neutral strategies are observed, and both strategies sell their power in this period and leave the partnership for a period of about 7 hours, but the Risk-neutral strategy. Due to the high demand, it participates in the market in the period of 7to14But the risk-averse continues to withdraw from the partnership due to conservatism until 6pm and resumes its participation at a high price, but with relatively less than the Risk-neutral.

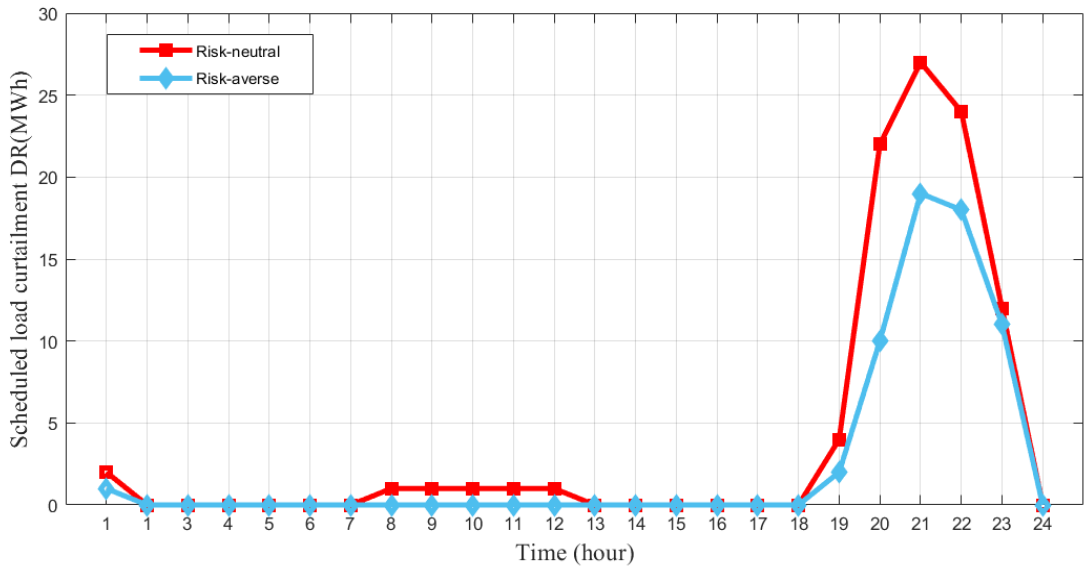

Fig 13. the scheduled load curtailment DRE

Fig 14. refers to the scheduled power of storage units. In this curve, consumption is related to the difference between the price of energy stored at low price and its sale at high price. This curve shows the combination of INT and DHM . Therefore, according to the analysis, Risk averse does not participate in the market for about 8 hours, but Risk neutral participates in the market for about 3 hours and sells some power, and is out of the market again and in the time periods of 8 to 15 and 18 to 24 again in the market.

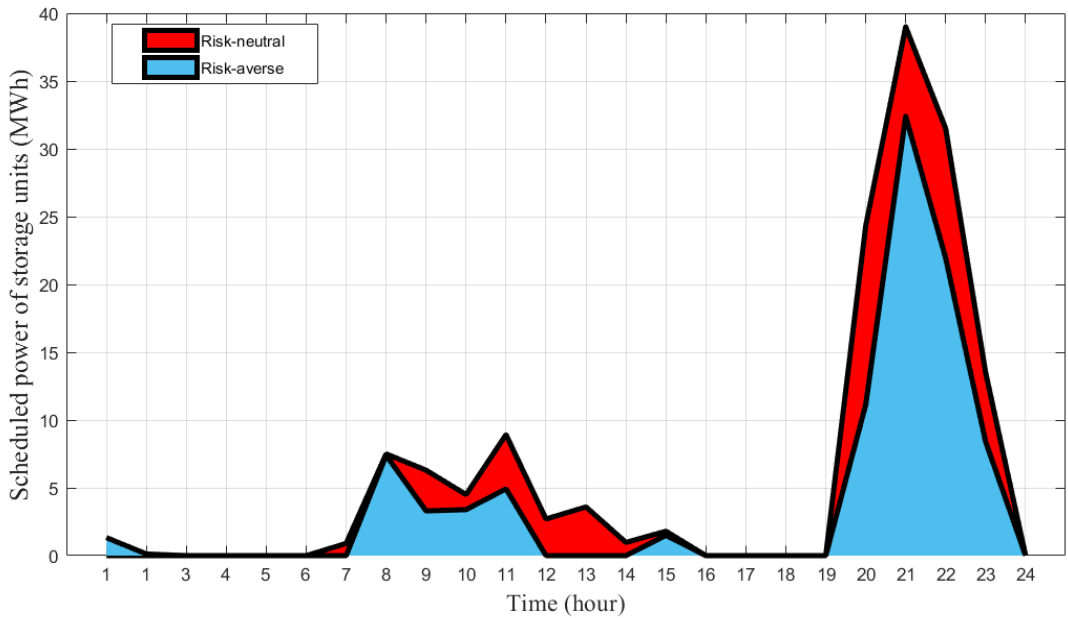

Fig 14. shows the scheduled power of storage units

Fig 15. represents the wind participation in the INT and DHMs. many energies, including wind, cannot be stored, and due to the unpredictable nature of the atmosphere, there are many changes in wind power. And wind production is independent of MP because it tries to sell the maximum wind produced in the markets, so the sales mode in the wind market has no effect on price, and all wind produced must be sold in DHM and INT the day before and in Real time compensates after the uncertainty and determination of wind production. Therefore, the power sold by wind is independent of price. So Risk-averse models the worst conditions and because one of the UCYs is related to wind, so the worst conditions for wind are reduced production and followed by a significant reduction in output according to the Risk-averse curve in the INT and DHM. In the INT, the amount of wind sold is less than DHM, which is due to the smaller volume of power in the INT. It is also observed in the INT that the fluctuation of the sold power is very high because it tries to make a profit from the price fluctuations. It is also observed that in the DHM these fluctuations are less due to the fact that the basic power that is predicted to sell in the DHM.

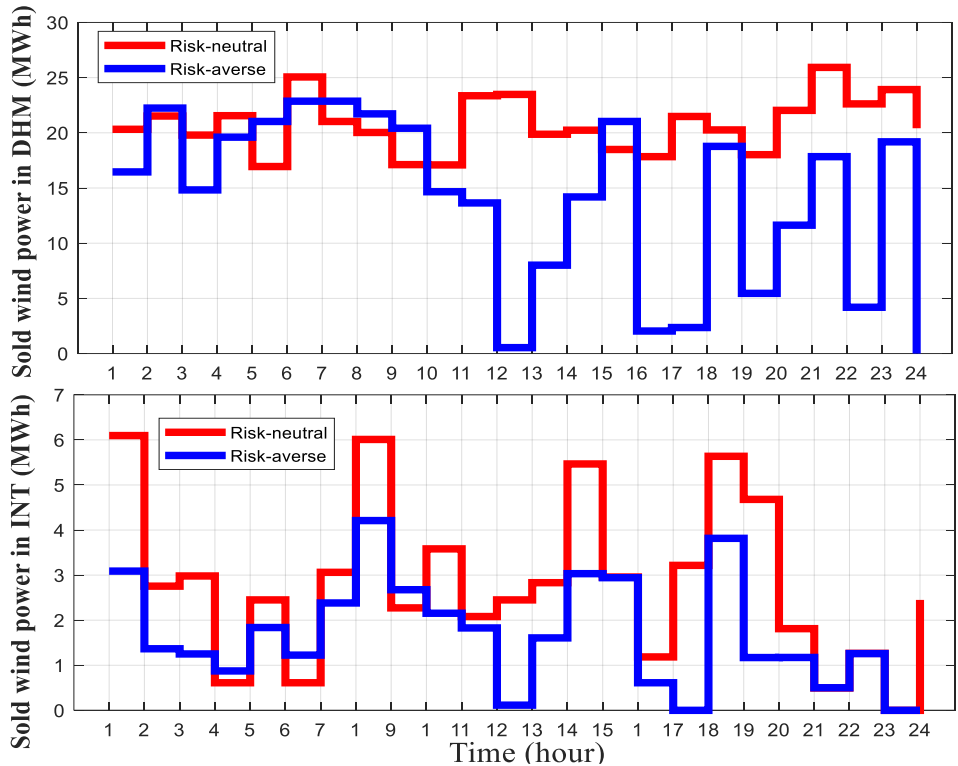

Fig 15.the wind participation in the INT and DHMs

In most of the advanced electricity markets in the world, the producers' bidding strategy is a combination of the bidding price and the amount of production that determines the share of the unit in the market. The price offer has a direct effect on the economic profit of the producers and does not only include the choice of the strategy adopted by the power plant itself, but also depends on the behavior of other producers and the operating conditions. Examining the issue of price quotation in a competitive environment is the most important issue that any business company considers. The profitability of companies also depends on how the price is offered, their risk management strategy and price fluctuations, the mechanisms of operation of the units and the rules governing the markets. The problem of price proposition is generally defined as finding the optimal price and determining the appropriate strategy to compete with other Examining the issue of price quotation in a competitive environment is the most important issue that any business company considers. The profitability of companies also depends on how the price is offered, their risk management strategy and price fluctuations, the mechanisms of operation of the units and the rules governing the markets. The problem of price proposition is generally defined as finding the optimal price and determining the appropriate strategy to compete with other sellers. At first, we will analyze the figure that related to offering hour 2. Analysis shows that both Risk-averse and Risk-neutral strategies have sold a volume of 21MW for up to $ 10. Due to its risk-taking, the Risk-neutral strategy sells 22 MW of its capacity for $ 19 and 24 MW of its capacity for $ 25 and then, it sold its power to 34MW at a price of $ 25 and increase its power to higher prices. Risk-averse strategy Unlike Risk-neutral strategy, due to its conservatism, at first it sells its power volume up to 23.5 MW at $ 10 and then increases its price to $ 25 and sells 24.5 MW of its power volume.

The next figure is related to offering at hour 5. The relationship between this curve and ONGs, and the existence of start-up costs instead of turning off the generator; (Due to increasing start-up costs), ONG prefers to sell 20MW of power at a lower price than the final cost, and then with the raising prices to $ 25, both the Risk-averse and Risk-neutral strategies will sell 20MW

of power. The figure related to the hour 7 offering in fig.10 represents that at first both the Risk-neutral and Risk-averse strategies sell the same amount of power equivalent to 20.5 MW at $ 25 price, then the Risk-averse strategy sells 27 MW at 25$ which is a higher amount compared to the Risk-neutral method due to being a conservative approach. Additionally, the Risk-neutral strategy sells 27MW and 29MW at the 33$ and 44$ MP. Furthermore, the offering related to hour 7 is also depicted in Fig. (10). According to the numerical findings of the investigation, the Risk-averse and Risk-neutral techniques sell the same amount of electricity at $25 MP at first. Then, the Risk-averse strategy tends to sell a relatively large amount of power equivalent to 27MW at the same $25 price due to being considered a conservative strategy. However, the Risk-neutral method waits for the price to increase in order to gain more profit by selling at higher prices as expected. Finally, as denoted in fig. (10) the Risk-neutral strategy sells 27 MW and 29MW at 33$ and 44$.As shown in fig.10 related to 3pm, At the price range of 18$ to 33$ the Risk-averse strategy sells 20.5 MW. However, compared to the Risk-averse strategy, the Risk-neutral strategy sells 21MW at the same price range. Furthermore, the Risk-neutral strategy sells 28MW power at the 50$ price. As mentioned before the Risk-averse strategy is a conservative approach. Thus, this strategy prefers to sell the most amount of its power at the peak hours. Consequently, at the 50$ the Risk-averse strategy due to 3pm being considered a peak hour strategy sells 48MW which is higher amount compared to the Risk-neutral strategy.

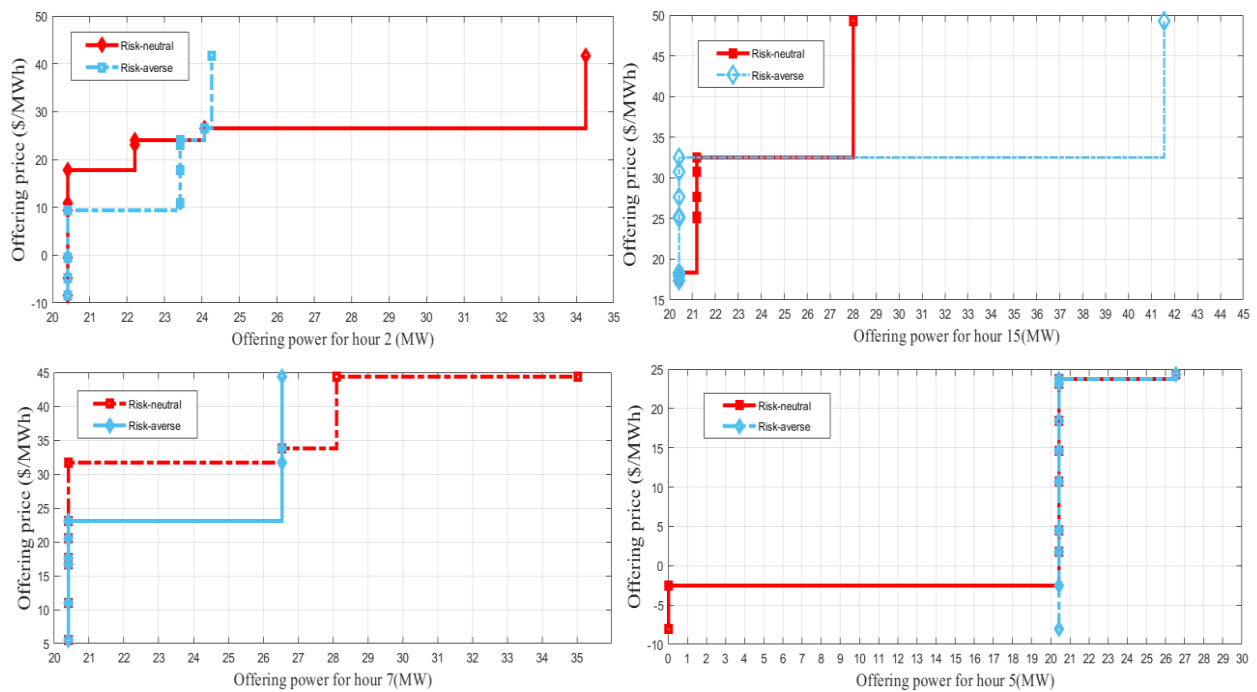

Fig 16. Offering in different hours of the day

## 7. conclusion

The uncertainty of VRP participation in the energy multimarket, including DA, ID, and RLT, was studied using a stochastic technique in this article. The P-robust technique is used to model the risk associated with the VRP's multi-market participation uncertainty. The p-robust strategy would reduce the complexity of the VRP's multi-market participation, resulting in faster market operations. Furthermore, wind generating uncertainty is seen as a serious difficulty since it causes RLT pricing fluctuations. As shown in the model, using the P-robust technique to manage the uncertainty and risk of the VRP leads in optimum decision making. Furthermore, combining the ONG, ENS, DRE, including load shifting and load curtailment, and the WPG might enhance market clearing, which is an appealing idea for further research. As represented in the paper, MRR and estimated profit in p=+ are 13.77% and $26, respectively. while in p=0.0747 they are reduced by 7.47% and $24410.06, respectively. As a result, the PHES' estimated profit and MRR are decreased by 6.81% and 45.75%, from p=+ to p=0.0747. As a result, it is evident that from p=+ to p=0.0747, a 6.81% reduction in estimated profit can result in a 45.75% reduction in MRR.